\documentclass[12pt]{article}
\usepackage{times}
\usepackage{geometry}
\usepackage[utf8]{inputenc}
\usepackage{enumitem,amssymb}
\usepackage{ragged2e}
\usepackage{graphicx}
\usepackage{sidecap}
\usepackage{natbib}
\usepackage{color}

\newlist{thematic}{itemize}{8}
\setlist[thematic]{label=$\square$}
\usepackage{pifont}
\newcommand{\cmark}{\ding{51}}%
\newcommand{\done}{\rlap{$\square$}{\raisebox{2pt}{\large\hspace{1pt}\cmark}}%
\hspace{-2.5pt}}

\newcommand\arcmin{\mbox{$^\prime$}}%
\newcommand\arcsec{\mbox{$^{\prime\prime}$}}%

\newcommand{\kms}         {km~s$^{-1}$}

\newcommand{\lcdm}        {$\Lambda$CDM}

\newcommand{\apj}{The Astrophysical Journal}
\newcommand{\apjl}{The Astrophysical Journal Letters}

\newcommand{\mnras}{Mon. Not. R. Astron. Soc.}
\newcommand{\aj}{Astron. J.}
\newcommand{\prl}{Physical Review Letters}
\newcommand{\araa}{Annual Rev. Astron. Astrophys.}
\newcommand{\aap}{Astron. Astrophys.}

\usepackage{aas_macros}

\begin{document}
\pagestyle{empty}
\raggedright
\huge
Astro2020 Science White Paper \linebreak

Dynamical Masses for a Complete Census of Local Dwarf Galaxies
 \linebreak
\normalsize

\noindent \textbf{Thematic Areas:} \hspace*{60pt} $\square$ Planetary
Systems \hspace*{10pt} $\square$ Star and Planet
Formation \hspace*{20pt}\linebreak $\square$ Formation and Evolution
of Compact Objects \hspace*{31pt} $\done$ Cosmology and Fundamental
Physics \linebreak $\square$ Stars and Stellar Evolution \hspace*{1pt}
$\done$ Resolved Stellar Populations and their
Environments \hspace*{40pt} \linebreak $\square$ Galaxy
Evolution \hspace*{45pt} $\square$ Multi-Messenger Astronomy and
Astrophysics \hspace*{65pt} \linebreak
  
\textbf{Principal Author:}

Name: Joshua D. Simon
 \linebreak						
Institution: Carnegie Observatories
 \linebreak
Email: jsimon@carnegiescience.edu
 \linebreak
%Phone: (626) 304-0256
% \linebreak
 
\textbf{Co-authors:} %(names and institutions)
%  \linebreak

Keith Bechtol (University of Wisconsin-Madison; kbechtol@wisc.edu)\linebreak
Alex Drlica-Wagner (Fermi National Accelerator Laboratory; kadrlica@fnal.gov)\linebreak
Marla Geha (Yale University; marla.geha@yale.edu)\linebreak
Vera Gluscevic (University of Florida, Princeton University; verag@princeton.edu)\linebreak
Alex Ji (Carnegie Observatories; aji@carnegiescience.edu)\linebreak
Evan Kirby (California Institute of Technology; enk@astro.caltech.edu)\linebreak
Ting S. Li (Fermi National Accelerator Laboratory; tingli@fnal.gov)\linebreak
Ethan O.\ Nadler (Stanford University; enadler@stanford.edu)\linebreak
Andrew B. Pace (Texas A\&M; apace@tamu.edu)\linebreak
Annika Peter (The Ohio State University; peter.33@osu.edu)\linebreak
Risa Wechsler (Stanford University; rwechsler@stanford.edu)\linebreak

\justify

\textbf{Abstract:} The 2020s are poised to continue the past two decades of significant advances based on observations of dwarf galaxies in the nearby universe.  Upcoming wide-field photometric surveys will probe substantially deeper than previous data sets, pushing the discovery frontier for new dwarf galaxies to fainter magnitudes, lower surface brightnesses, and larger distances.  These dwarfs will be compelling targets for testing models of galaxy formation and cosmology, including the properties of dark matter and possible modifications to gravity.  However, most of the science that can be extracted from nearby dwarf galaxies relies on spectroscopy with large telescopes.  We suggest that maximizing the scientific impact of near-future imaging surveys will require both major spectroscopic surveys on $6-10$~m telescopes and multiplexed spectroscopy with even larger apertures.

\pagebreak
\pagestyle{plain}
\setcounter{page}{1}
%\justify

\section{Dwarf Galaxy Discovery in the Next Decade}

The past 15 years have been a golden age for the study of dwarf
galaxies, particularly at the faintest luminosities.  Each of the
past two decades has seen the discovery of at least as many new Milky Way
satellite galaxies as were known at the start of the decade, and the
2020s are poised to continue this trend.  In this contribution we
describe the scientific opportunities that will be enabled by upcoming
searches for new dwarf galaxies in the nearby universe and follow-up
spectroscopy of those objects.  Specifically, major advances will be
possible in measuring the faint end of the galaxy luminosity function,
establishing the lower limit for galaxy formation, constraining the
stellar mass-halo mass relation in the dwarf galaxy regime,
investigating the quenching of dwarf galaxies, and testing dark matter models via indirect
detection experiments and other astrophysical measurements.  Several
complementary white papers focus on the utility of dwarfs for constraining dark matter physics (Li et al.,
Drlica-Wagner et al., Gluscevic et al., Simon et al.) and on the stellar populations and
chemical abundances of nearby dwarfs (Weisz \& Boylan-Kolchin, Ji et
al.).

The modern era of dwarf galaxy discovery was enabled by the first
wide-field digital sky surveys.  As a result of ongoing surveys,
within the next $\sim1-2$ years the full high-latitude sky will have
been surveyed to a depth of $g \sim 23$ \citep[e.g.,][]{DrlicaWagner2016,Dey2019}.  Dwarf galaxy searches using
these data rely on a standard matched-filter approach, identifying
overdensities of stars consistent with an old, metal-poor population
\citep[e.g.,][]{willman2005,belokurov2006,bechtol2015,koposov2015}.
Existing data are sensitive to stellar systems as faint as M$_{\rm V}
\approx -6$ out to distances of $\sim400$~kpc, M$_{\rm V} \approx -4$ out to
distances of $\sim200$~kpc, and M$_{\rm V} \approx -2$ out to distances of
$\sim50$~kpc \citep{koposov2008,walsh2009,bechtol2015,Simon2019}.

In the next decade, LSST will provide dramatically deeper imaging,
enabling the discovery of M$_{\rm V} = 0$ dwarf galaxies (comparable
to the faintest objects currently known) anywhere within the virial
radius of the Milky Way, and M$_{\rm V} = -6$ dwarfs throughout the
Local Group.  Beyond the Local Group, dwarf galaxies brighter than
M$_{\rm V} \sim -8.5$, similar to the classical satellites of the
Milky Way, will be detectable as overdensities of resolved stars out
to distances of $\sim5$~Mpc.

\vspace{-0.5cm}
\subsection{Milky Way satellites}
\vspace{-0.2cm}

The ultra-faint satellite galaxies of the Milky Way are the least luminous, most chemically pristine, and most ancient galaxies known. As some of the most extreme examples of galaxy formation, ultra-faint dwarfs uniquely probe galaxy formation, the origin of the heavy elements,  and the fundamental physics of dark matter \citep[e.g.,][and references therein]{Simon2019}.  

The least luminous galaxies discovered thus far contain only a few hundred stars and have been found exclusively in the inner regions of the Milky Way because of observational selection effects. Although the known population of Milky Way dwarf galaxies has grown from $\sim 25$ to more than 50 in recent years \citep[e.g.,][]{bechtol2015, koposov2015, DrlicaWagner2015}, our current census is certainly incomplete.
For example, the HSC-SSP collaboration has detected two ultra-faint galaxy candidates in the first 300~deg$^2$ of that survey \citep{Homma2016,Homma18}; these objects would be undetectable in all previous wide-field surveys. HSC is shallower than the depth that will be achieved by LSST over half the sky---an area 60 times larger than the current HSC-SSP footprint. Thus, based on the results of SDSS, HSC, DES, etc., LSST is predicted to detect tens to hundreds of new ultra-faint Milky Way satellites, mainly at larger distances and fainter luminosities than those accessible with current-generation surveys \citep{koposov2008,Tollerud:2008,Hargis:2014,Newton:2018,Jethwa:2018,Kim:2017iwr,Nadler:2018}. 
In addition, novel techniques, such as the use of the correlated phase space motions of stars \citep{1507.04353,1805.02588,Torrealba18} or clustering of variable stars \citep{1507.00734} could further expand the sample of new dwarfs.
Over the coming decade, these discoveries will greatly enhance our understanding of the faintest galaxies, providing a unique testing ground for our understanding of galaxy formation and fundamental physics.

\vspace{-0.5cm}
\subsection{Local Group dwarfs}
\vspace{-0.2cm}

In the past two decades, dwarf galaxies in the Local Group but beyond the gravitational influence of the Milky Way have been discovered exclusively in the vicinity of M31 \citep[e.g.,][]{Mcconnachie2008,Richardson2011}.  
The near-future and proposed wide-area optical and near-IR imaging surveys of LSST, Euclid, and WFIRST will make it possible to detect 
ultra-faint dwarf galaxies throughout the entire Local Group (and in the nearby field) for the first time.  
This unbiased census of the faintest galaxies in a broad range of environments is essential for revealing the effects of environment on their formation and evolution.  In particular, finding ultra-faint dwarfs beyond the virial radius of a massive host is key to understanding whether these systems are quenched solely by reionization or in combination with environmental or internal processes \citep{Brown2014,Fillingham2018,RodriguezWimberly2019}.

\vspace{-0.5cm}
\subsection{Local universe dwarfs}
\vspace{-0.2cm}
Recent searches for dwarf galaxies beyond the Local Group have so far focused on finding satellites around Local Group analogs.   Resolved stellar searches of nearby galaxies ($< 5$\,Mpc) have now produced satellite luminosity functions for several neighboring systems, including M81, Centaurus~A, and NGC~2403 \citep[e.g.,][]{Chiboucas2009,Carlin2016}, enabling comparisons with the faint end of the luminosity function of the Milky Way and M31 \citep{Smercina2018,Crnojevic2019}.   
Several isolated dwarfs with M$_{\rm V} \sim -9$ have also recently been discovered serendipitously in HST imaging at distances up to 10~Mpc \citep{Makarova18,Martinez-Delgado2018,Bedin2019}.
Between $\sim5 - 20$\,Mpc, searches for unresolved low surface brightness features have successfully detected low-luminosity dwarf galaxies \citep{Danieli2018}; as with closer dwarfs, follow-up observations (spectroscopy or HST imaging) are required to ensure pure samples \citep[e.g.,][]{Danieli2017}.   Between $20-50$\,Mpc, dwarf galaxies are difficult to distinguish from more distant galaxies via photometry alone.  Progress is being made to develop reliable photometric redshifts and/or efficient photometric cuts, but follow-up spectroscopy is required in this case as well \citep{Geha2017}.  In all three distance regimes, searches are expensive and necessarily focused on a particular environment (e.g., satellites of Milky Way-like systems) due to limited resources.

The next generation of imaging surveys will facilitate dwarf galaxy detection over a wide range of environments beyond the Local Group.   LSST and other deep imaging surveys will expand the sky coverage and volume over which the above dwarf-finding techniques can be applied.  In particular, the wide-area coverage of LSST will allow detection of dwarf galaxies in more isolated regions where galaxy processes related to environment are minimized.   Field galaxies at these luminosities
provide critical control samples from which to isolate environmental galaxy formation processes \citep[e.g.,][]{Dickey2019}, as well as enabling new tests of the missing satellite and too big to fail problems \citep[e.g.,][]{GarrisonKimmel2017,Read2017}.

\section{Spectroscopy of Newly Identified Dwarf Galaxies}

\vspace{-0.2cm}
\subsection{Confirmation of dwarf galaxy candidates}
\vspace{-0.2cm}

Imaging surveys discover dwarf galaxy \emph{candidates}.  The brightest such
objects are obviously real systems, and depending on their
luminosities and sizes some can be immediately classified as dwarf
galaxies without follow-up information.  However, as surveys push toward lower
luminosities and larger distances, spectroscopic follow-up
becomes more critical, not only to measure their physical properties,
but also simply to confirm which systems are dwarfs and which are not.

The primary confounding objects for dwarf galaxy searches are faint, extended star clusters.  Dwarf galaxies differ from star clusters in that they form and still reside in dark matter halos.  Observationally, the presence of dark matter manifests itself as either an inferred dynamical mass much larger than the stellar mass, or a spread in heavy element content indicating the retention of supernova ejecta \citep{ws2012}.  Dwarf galaxy candidates can therefore be confirmed by detecting a significant velocity or metallicity dispersion, which can both be measured from medium resolution ($R\gtrsim6,000$) spectroscopy of individual member stars.
The lowest-luminosity galaxies have typical velocity dispersions of ${\sim}3$~\kms\ and metallicity dispersions of ${\sim} 0.2$ dex \citep{Simon2019}.  Accurate velocity dispersion measurements require velocity measurements with ${\sim}1$~\kms\ accuracy for a sample of ${\sim}20$ stars.  Metallicity dispersions are usually detected after velocity dispersions have been resolved, but in some cases they constitute the primary evidence for the dwarf galaxy nature of a system \citep[e.g.,][]{Kirby2013}.  At the lowest surface brightnesses, random groupings of foreground stars may also become significant contaminants. These are easily distinguished from true dwarf galaxies with spectroscopy.

\vspace{-0.5cm}
\subsection{The dwarf galaxy dynamical mass function}
\vspace{-0.2cm}

Once an object is confirmed as a dwarf galaxy, one of the primary properties of interest is its dynamical mass.
Dynamical masses directly trace the nonlinear small scale matter power spectrum,
set the normalization for indirect dark matter detection experiments through $J$-factors,
and determine which halos in galaxy formation simulations should be compared to which observations.

In dwarf spheroidal galaxies, dynamical masses are determined by measuring line-of-sight stellar velocities and applying a mass estimator that assumes the stars are in dynamical equilibrium. 
This is most difficult for faint and/or far away dwarf galaxies, where only a handful of stars are bright enough for velocity measurements.
For these galaxies, one typically must assume that all stars in a galaxy belong to a single equilibrium component, which can be characterized by a velocity dispersion and half-light radius \citep{Wolf:2009tu}.
However, velocity dispersions can be inflated by stars in binary systems, which require multi-epoch observations to detect, and the assumption of dynamical equilibrium can be violated if a galaxy has recently experienced significant tidal forces.  \emph{We emphasize that nearly all scientific uses of nearby dwarf galaxies, ranging from comparison with numerical simulations to searches for gamma-ray emission from dark matter annihilation, rely fundamentally on stellar spectroscopy.}

\vspace{-0.5cm}
\subsection{Nature of dark matter}
\vspace{-0.2cm}

\noindent \emph{Indirect dark matter detection.} The satellite galaxies of the Milky Way are especially promising targets
for the indirect detection of dark matter annihilation due to their large dark matter content, low Galactic foregrounds, and lack of conventional astrophysical mechanisms for producing energetic standard model particles.
The increasing population of ultra-faint satellites will further inform studies of fundamental physics through correlated observations with X-ray and $\gamma$-ray telescopes.

\begin{SCfigure}
  \includegraphics[width=0.6\textwidth]{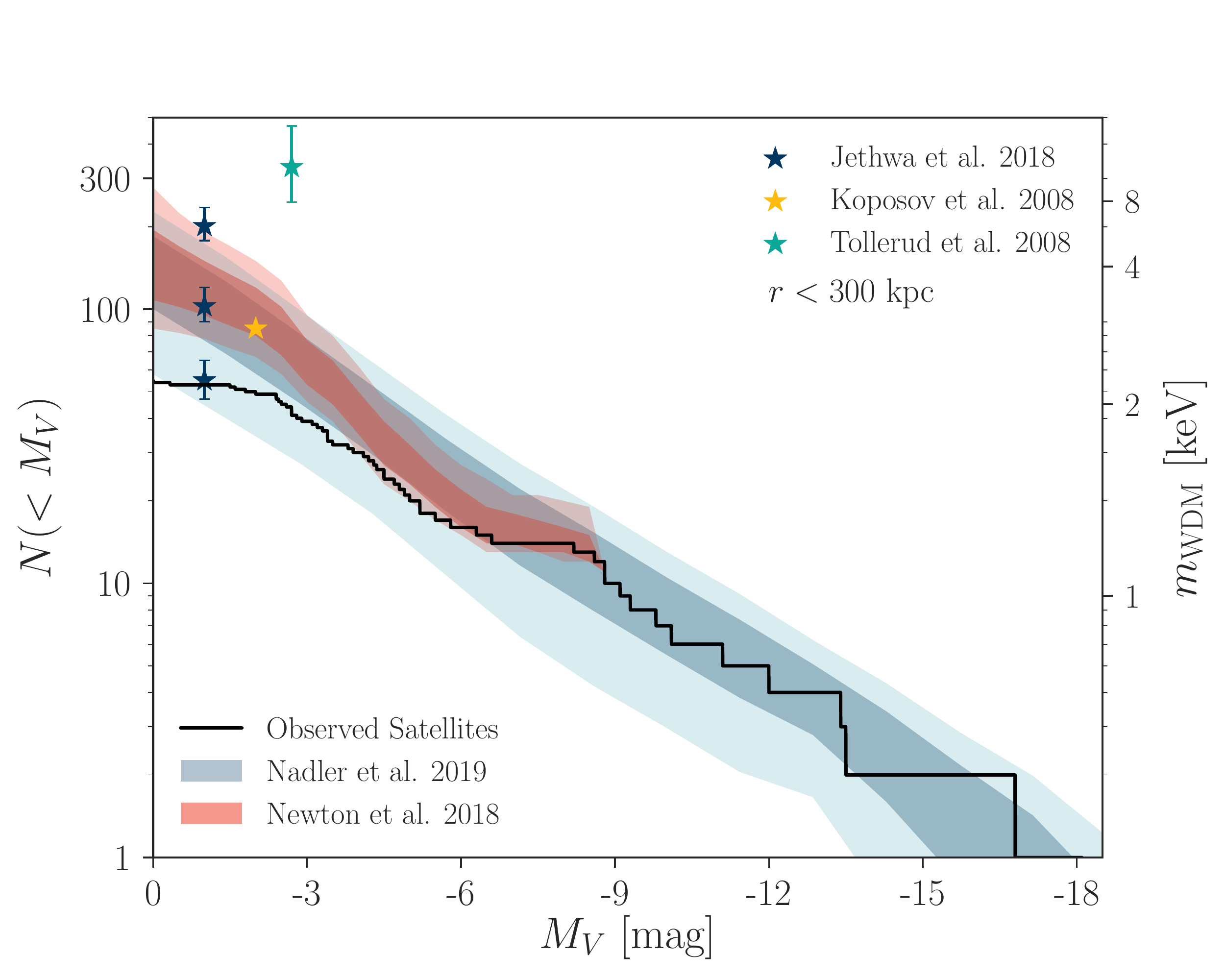}
  \caption{Literature comparison of the total predicted number of satellites within 300~kpc of the MW inferred from classical and SDSS-discovered Milky Way satellites. These predictions are compared to all observed satellites (including candidate systems; black line). The right-hand axis shows the number of satellites above $10^{6}$~M$_{\odot}$ at $z=9.3$ as a function of WDM particle mass \citep{Kim:2017iwr}. Figure adapted from \cite{Nadler:2018}.}
  \label{fig:lsst_predictions}
\vspace{-0.5cm}
\end{SCfigure}

\noindent \emph{Small scale power.} Many interesting dark matter scenarios (warm dark matter, fuzzy dark matter, many types of self-interacting dark matter, and dark matter-baryon scattering) suppress the small-scale power spectrum, which manifests as an under-abundance of dwarf galaxies in the lowest-mass dark matter halos relative to \lcdm\ predictions.
In the coming decade, observations of Milky Way satellites offer an exciting testing ground for these models of particle dark matter.
The measured abundance, luminosity function, and radial distribution of Milky Way satellites \emph{already} place competitive constraints on warm dark matter (WDM) particle mass at the level of 3--4 keV \citep[e.g.,][see Fig.~\ref{fig:lsst_predictions}]{Jethwa:2018,Kim:2017iwr}.
In the 2020s, this sensitivity may be improved to WDM particle masses of ${\sim} 17$ keV, thereby testing popular models of particle dark matter invoked to solve small-scale challenges to $\Lambda$CDM \citep{Bullock:2017}.
Similarly, the current population of Milky Way dwarfs was recently used to derive some of the most stringent astrophysical constraints on dark matter-proton interactions, far surpassing the sensitivity of cosmological probes (Nadler, Gluscevic, \& Boddy, in prep.).  Future dwarf galaxy discoveries will further tighten these limits.

\vspace{-0.2cm}
\section{Spectroscopic Resource Needs}
\vspace{-0.2cm}

Although next-generation imaging surveys are necessary for discovery, most of the science related to nearby dwarf galaxies requires spectroscopy.  As discussed above, confirmation of dwarf candidates and mass and metallicity measurements rely on spectroscopic follow-up observations.  Even for dwarf galaxies identified in the past two decades, which are generally brighter and closer than those that will be found in the 2020s, nearly all spectroscopy has been carried out with large ($6-10$~m) telescopes.  Increasing time investments on these and even larger facilities will be necessary for the discoveries to come (see below).  {\color{blue} \emph{We recommend that both extremely large telescopes and new dedicated spectroscopic survey facilities on large ($>6$~m) telescopes be pursued in order to facilitate a wide range of dwarf galaxy science.}}

\begin{figure}
  \centering
  \includegraphics[width=0.49\textwidth]{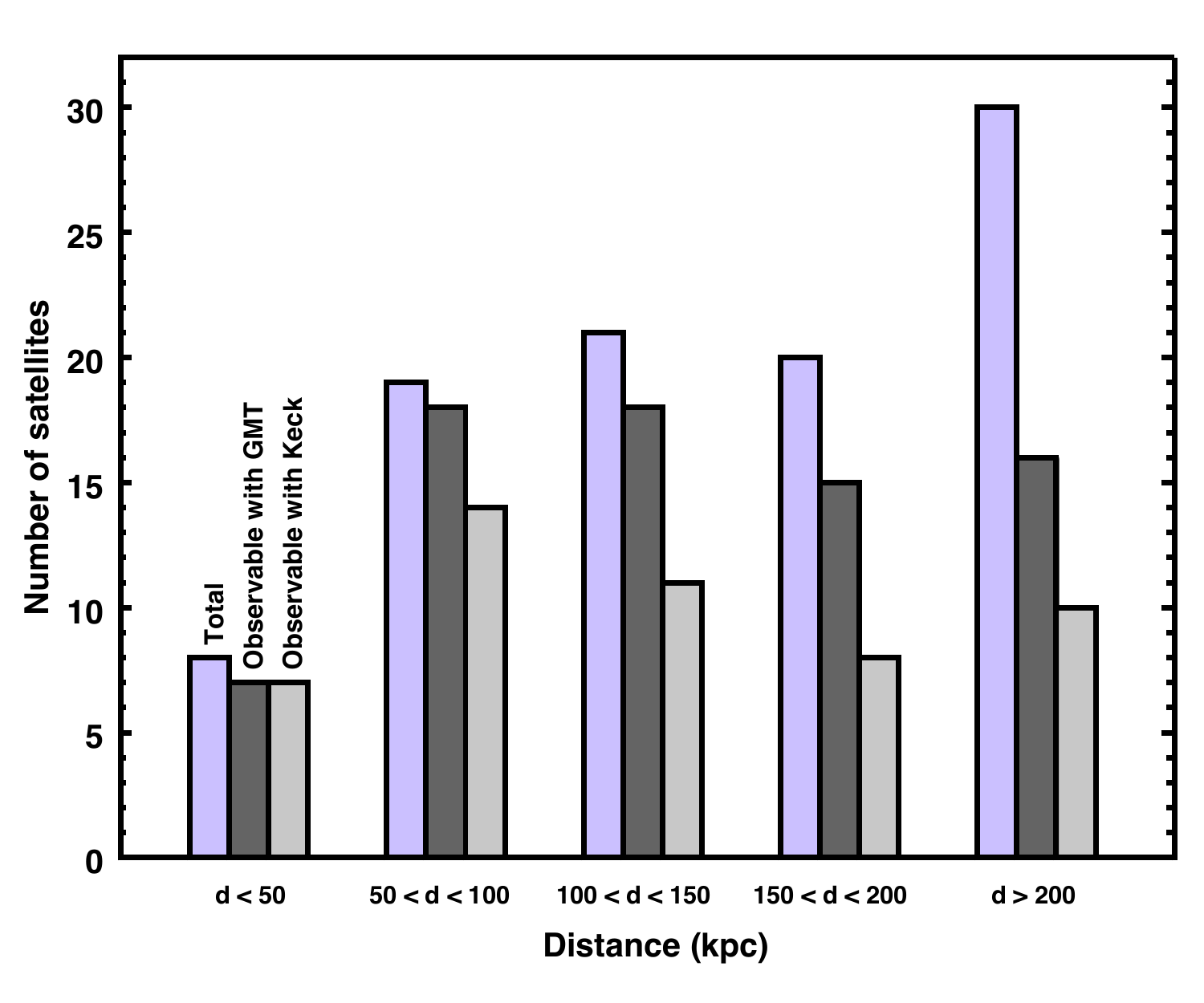}
  \includegraphics[width=0.50\textwidth]{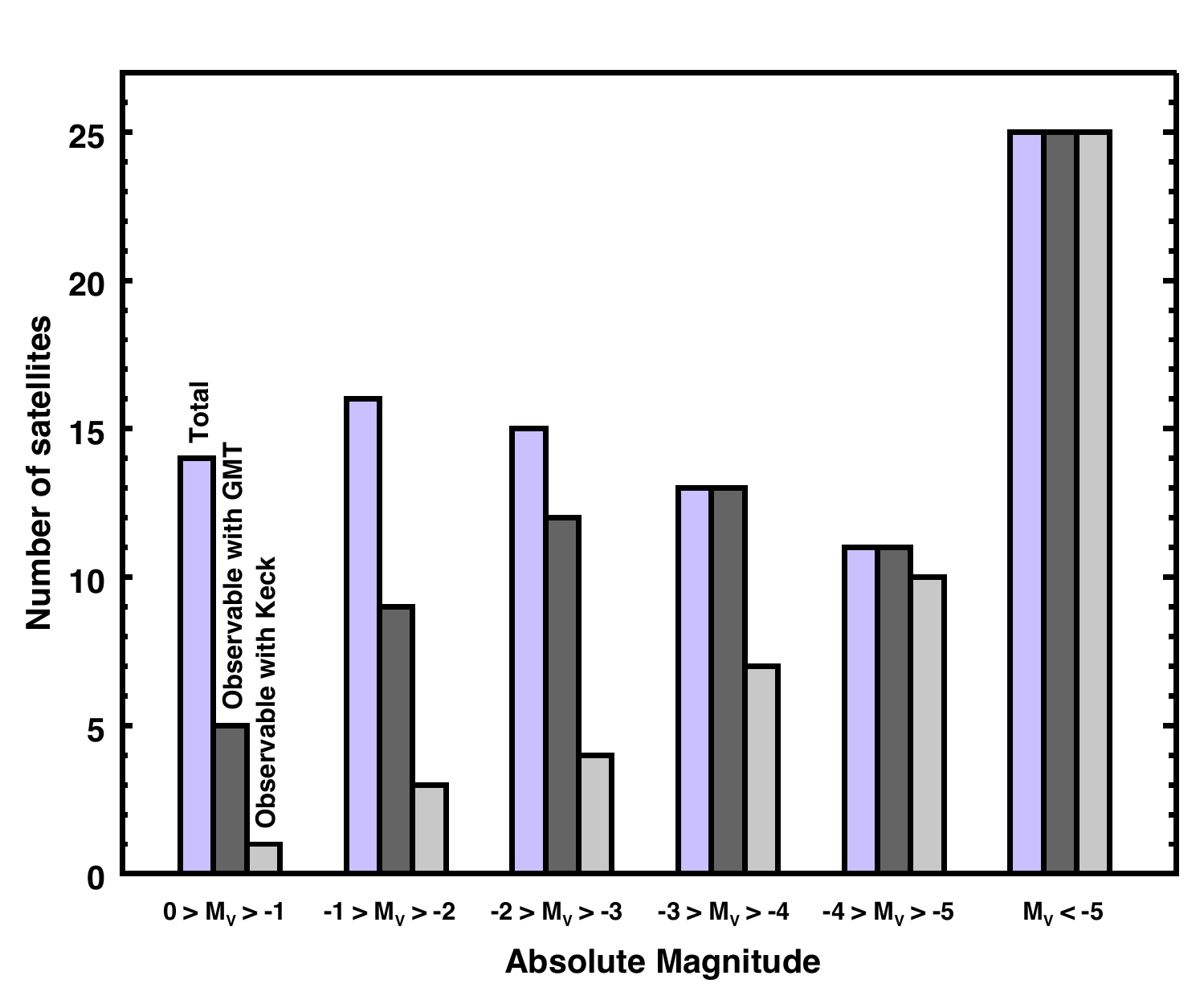}
  \caption{Limits on spectroscopic follow-up capabilities for the LSST population of Milky Way satellites as a function of distance (left) and magnitude (right). Current telescopes will be able to measure velocity dispersions for $\sim50\%$ of the expected satellites, while an ELT can measure velocity dispersions for $\sim80\%$.  Figure from \citet{DrlicaWagner2019}.}
  \label{fig:specfollowup_distance}
\vspace{-0.4cm}
\end{figure}

{\bf Estimated needs:} Several previous studies have estimated the spectroscopic resources that will be needed to follow-up discoveries from LSST and other surveys.  \citet{najita2016} concluded that mass measurements for new Milky Way satellites alone will add up to 3200~hours of spectroscopy on an $8-10$~m telescope; i.e., \emph{these observations would require more than a year of dedicated time on one of the largest current telescopes.}  Comparable data could be acquired $\sim10\times$ more rapidly with an extremely large telescope (ELT).  Of course, obtaining spectroscopy for more distant dwarfs will be even more time-consuming, and we understand that there may be interest in observing non-dwarf galaxies with these facilities as well.
Using updated projections for the expected satellite population, \citet{DrlicaWagner2019} showed that a significant fraction of dwarfs that will be found by LSST are too faint or too distant for follow-up with existing facilities (Fig.~\ref{fig:specfollowup_distance}).  Under the assumption that accurate velocity measurements are needed for 20 member stars in a dwarf galaxy to determine its velocity dispersion, they calculated that masses for only 50\% of the LSST satellite population can be successfully measured with $8-10$~m telescopes.  In contrast, ELT spectroscopy could measure dispersions for 80\% of the anticipated satellites.  

{\bf Required spectrograph characteristics:} Dwarf galaxies in the local universe have typical sizes ranging from tens of arcseconds at distances of a few Mpc to $>1^{\circ}$ for the most extended Milky Way satellites.  A minimum field of view of $\sim50$~arcmin$^{2}$ is needed to efficiently observe most systems, and a field of view $\gtrsim30\arcmin$ diameter is highly advantageous for the closest dwarfs.  Many recently identified dwarfs have velocity dispersions of $\lesssim3$~\kms\ \citep[e.g.,][]{Koposov2018,Li2018}, indicating that velocity measurements at $\sim1$~\kms\ accuracy or better will be needed for accurate characterization.  Results from existing instruments suggest that spectral resolution $R>6000$ is necessary to achieve this accuracy \citep[e.g.,][]{Koposov2011,Kirby2015,Simon2017}.  Newly discovered dwarfs are likely to contain $\lesssim200$ stars bright enough for spectroscopic follow-up, so multiplexing is less of a design driver for dwarf galaxy spectroscopy than for other science cases.  More important is the target density that can be accommodated, because stars 5\arcsec\ or less apart often need to be observed simultaneously.  We recommend that next-generation instruments be designed with the necessary field of view, velocity stability, and target spacing to enable efficient spectroscopy of dwarf galaxies in the nearby universe.

\pagebreak
\bibliographystyle{apj}
%\bibliography{dwarf_refs,main}

\end{document}